\documentstyle[12pt,aasms4]{article}
\received{}
\revised{}
\accepted{}
\cpright{}{}\journalid{}{}
\articleid{}{}
\paperid{}
\slugcomment{Submitted to the Astrophysical Journal}
\lefthead{Kwitter \& Henry}
\righthead{C Abundances in PNe}
\begin{document}
\title{A New Look At Carbon Abundances in Planetary Nebulae. 
III. DDDM1, IC~3568, IC~4593, NGC~6210, NGC~6720, 
NGC~6826, \& NGC~7009}
\author{K.B. Kwitter\footnote{Visiting Astronomer, Kitt Peak 
National Observatory, National Optical Astronomy Observatories, 
which is operated by the Association of Universities for Research in 
Astronomy, Inc. (AURA) under cooperative agreement with the 
National Science Foundation.}}
\affil{Department of Astronomy, Williams College, Williamstown, 
MA
01267; kkwitter@williams.edu}
\and
\author{R.B.C. Henry$^1$}
\affil{Department of Physics \& Astronomy, University of 
Oklahoma, Norman, OK  73019; henry@phyast.nhn.uoknor.edu}

\begin{abstract}
This paper is the third in a series reporting on a study of carbon
abundances in a carefully chosen sample of planetary nebulae
representing a large range in progenitor mass and metallicity.  We
make use of the IUE Final Archive database containing
consistently-reduced spectra to measure line strengths of C~III]
$\lambda$1909 along with numerous other UV lines for the planetary
nebulae DDDM1, IC~3568, IC~4593, NGC~6210, NGC~6720, NGC~6826, \&
NGC~7009.  We combine the IUE data with line strengths from optical
spectra obtained specifically to match the IUE slit positions as
closely as possible, to determine values for the abundance ratios
He/H, O/H, C/O, N/O, and Ne/O.  The ratio of C~III] $\lambda$1909/C~II
$\lambda$4267 is found to be effective for merging UV and optical
spectra when He~II $\lambda$1640/$\lambda$4686 is unavailable. Our
abundance determination method includes a 5-level program whose
results are fine-tuned by corrections derived from detailed
photoionization models constrained by the same set of emission
lines. All objects appear to have subsolar levels of O/H, and all but
one show N/O levels above solar.  In addition, the seven planetary
nebulae span a broad range in C/O values.  We infer that many of our
objects are matter bounded, and thus the standard ionization
correction factor for N/O may be inappropriate for these PNe.
Finally, we estimate C/O using both collisionally-excited and
recombination lines associated with C$^{+2}$ and find the
well established result that abundances from recombination lines
usually exceed those from collisionally-excited lines by several
times.
\end{abstract}


\section{Introduction}

We report further on a project whose aim is to determine the stellar
yield of carbon as a function of stellar mass and metallicity for
intermediate-mass stars, those in the mass range 0.8 $<M<$
8$M_{\odot}$.  These stars are predicted to produce as much as 50\% of
the carbon in the Galaxy (Henry, Kwitter, \& Buell 1998).  In our
general study, measured line intensities are used in 5-level atom and
photoionization model calculations to determine the abundance ratio of
C/O in particular, but also of He/H, O/H, N/O, and Ne/O for 22
planetary nebulae (PNe) selected to represent a broad range in
progenitor mass and metallicity.  Ultimately, we will use our
abundance results as constraints for our own stellar evolution models
in order to derive the stellar yield of carbon as a function of these
two parameters.  We have measured IUE spectra of PNe containing
strong, collisionally-excited carbon emission lines, spectra which
have been re-reduced in a systematic way and are now in the Final
Archive.  In two earlier papers, the UV data were joined with optical
data from the literature; beginning with this paper, we use new
optical data acquired specifically for this project.

In our first paper (Henry, Kwitter, \& Howard 1996; hereafter
Paper~I)  we listed our sample objects, described our project in 
detail, and presented results for the first four PNe.  In our second 
paper (Kwitter \& Henry 1996; hereafter Paper~II) we reported on 
five additional PNe. In the current paper we describe our analysis 
of seven more objects: DDDM1, IC~3568, IC~4593, NGC~6210, 
NGC~6720, NGC~6826, \& NGC~7009.  Future papers will report 
on the remaining six objects, as well as present and discuss stellar 
model predictions of carbon yields for intermediate-mass stars. 
Section {\S\ }2 describes the data used in the analysis of these 
seven objects; the abundance calculations and results are presented 
in {\S\ }3; and a summary is contained in {\S\ }4.  More detailed
discussion of the project and procedures can be found in Paper~I.

\section{The Data}

\subsection{Optical Observations}

The optical data were obtained at KPNO during 18-21 May 1996 with the
2.1m Goldcam CCD spectrometer.  The chip is a Ford 3K $\times$ 1K CCD
with 15$\mu$ pixels. We used a 5$\arcsec$- wide slit that extended
285$\arcsec$ in the E-W direction, with a spatial scale of
0$\farcs$78/pixel. Using a combination of two gratings, we obtained
spectral coverage from 3700-9600\AA\ with overlapping coverage from
$\sim$5750 - 6750\AA.  Wavelength dispersion was 1.5 \AA/pixel
($\sim$8 \AA\ FWHM resolution) for the blue, 1.9 \AA/pixel ($\sim$10
\AA\ FWHM resolution) for the red. The usual bias and twilight
flat-field frames were obtained each night, along with HeNeAr
comparison spectra for wavelength calibration and standard star
spectra for sensitivity calibration.  Since the chip is thinned, it
produces interference fringes visible in the red.  In our red spectra
the fringes appear at the $\pm$1\% level at $\sim$7500\AA\ and
increase in amplitude with increasing wavelength: $\pm$1.5\% at
8000\AA, $\pm$4.5\% at 8500\AA, $\pm$6\% at
9000\AA. However, even at their worst, {\it i.e.\/}, at
$\sim$$\lambda$9500, the longest wavelength we measure, the fringe
amplitude reaches only about $\pm$7\%. Internal quartz flats
were taken at the position of each object both before and after the
object integrations in anticipation of removing the fringes during
data reduction. It turned out, however, that more noise was introduced
in this process than was removed; we therefore decided to leave the
fringes untouched, and to accept this additional uncertainty in our
line intensities longward of $\sim$7500\AA.

The original images were reduced in the standard fashion using
IRAF\footnote{IRAF is distributed by the National Optical Astronomy
Observatories, which is operated by the Association of Universities
for Research in Astronomy, Inc. (AURA) under cooperative agreement
with the National Science Foundation.}. Using tasks in the {\it kpnoslit\/} package, these
two-dimensional spectra were converted to one dimension by
extracting a specific section along the slit. The location of the
extracted section was chosen to maximize the overlap with the IUE
slit.

\subsection{UV Data}

All UV spectra used for this project have been obtained from the IUE
Final Archive.  Spectra in the Final Archive have been systematically
and uniformly re-processed by IUE staff using the NEWSIPS algorithms,
and represent the best available calibration of these data.  The
spectra we use are all short-wavelength (SWP), low-dispersion, and
large-aperture (21$\farcs$7$\times$9$\farcs$1) spectra. Table 1 lists
the spectra that were measured for each of the seven program objects
considered here, along with their intergation times.

\subsection{Slit Positions}

The placement of the Goldcam slit in each target PN was chosen to 
coincide as closely as possible with the location of the best IUE 
observations for which detailed positional information was 
available. Since the position angle of the Goldcam slit is fixed at 
90$\arcdeg$ while the IUE slit position angle is not, the quality of 
the overlap varies and will be described below for each object. We 
also note that because of the  2:1 relative slit widths, the largest 
possible overlap of the Goldcam slit onto the IUE slit is 
$\sim$50\%. 

 For each object we now describe the IUE and optical
observations with regard to slit position. In general, our optical N-S
offsets from the central star or the center of the nebula match the
IUE N-S offsets; the E-W component (if any) of the IUE offset is
matched in the extraction process that creates a one-dimensional
spectrum from the appropriate portion of the two-dimensional spectrum.

{\it DDDM1}:  All three of the IUE spectra were centered on the 
central star, as were our optical spectra. The position angles of the 
IUE spectra are very similar: 349$\arcdeg$, 346$\arcdeg$, 
and 345$\arcdeg$. Since the optical diameter of DDDM1 is 
0$\farcs$6 (Acker et al. 1992), the entire nebula was included in 
both IUE and optical observations, rendering moot the issue of 
overlap.

{\it IC~4593}:  Both IUE spectra were centered 7$\arcsec$~S of the 
center of light (which we are assuming to coincide with the center 
of the nebula), at a position angle of 120$\arcdeg$. Our slit was 
centered 3$\arcsec$~S of the nebula center, which we judge to be 
fair overlap.

{\it IC~3568}:  All three IUE spectra were centered on the central 
star at position angles of 13$\fdg$6, 108$\arcdeg$, and  
328$\arcdeg$. Our slit was centered 4$\arcsec$~N of the central 
star, and we judge the overlap to be reasonably good.

{\it NGC~7009}:  Positions for the four IUE spectra we used for 
NGC~7009 were kindly provided by Dr. F. Bruhweiler (private 
communication). Position {\it a} was located 9$\farcs$3~S, 
9$\arcsec$~E of the central star, at a position angle of 
139$\arcdeg$.  For position {\it b}, the IUE slit was 9$\farcs$3~S, 
3$\farcs$5~W, again at position angle 139$\arcdeg$. To cover both 
of these IUE positions, the Goldcam slit was located 9$\arcsec$~S. 
Position {\it c} is located 4$\arcsec$~N, 22$\arcsec$~E, at a 
position angle of 334$\arcdeg$ and includes the east ansa. Position 
{\it d} is 4$\arcsec$~N, 8$\arcsec$~E at a position angle of 
139$\arcdeg$. For both of these positions, the Goldcam slit was 
placed 4$\arcsec$~N. For all four positions in NGC~7009, the 
overlap is reasonably good.

{\it NGC~6720}: Position {\it a} is located 11$\farcs$2~S, 
16$\farcs$6~E of the central star at position angle 124$\arcdeg$. 
The Goldcam slit was placed 11$\farcs$2~S, yielding reasonably 
good overlap. Position {\it b} is 16$\arcsec$~N, 42$\farcs$8~E at 
position angle 295$\arcdeg$. The Goldcam slit was 17$\farcs$3~N, 
giving reasonably good overlap.

{\it NGC~6210}: Both IUE spectra were positioned 4$\arcsec$~N, 
8$\arcsec$~E at position angle 10$\fdg5$. The Goldcam slit was 
centered 4$\arcsec$~N, producing fair overlap.

{\it NGC~6826}: Position {\it a} is 10$\arcsec$~S, 10$\arcsec$~W 
of the central star at position angle 218$\arcdeg$. Position {\it b} is 
9$\arcsec$~S, 5$\arcsec$~E at position angle 58$\arcdeg$. For 
both of these positions, the Goldcam slit was placed 9$\arcsec$~S, 
and the resulting overlap was good. Position {\it c} is 
10$\arcsec$~N, 10$\arcsec$~E at position angle 218$\arcdeg$. The 
Goldcam slit was 10$\arcsec$~N, with good overlap.

\subsection{Line Strengths}

Strengths of all optical and UV lines were measured using {\it
splot} in IRAF and are reported in Table~2.
Fluxes uncorrected for reddening are presented in
columns labelled F($\lambda$), where these flux values have
been normalized to H$\beta$=100 using our observed value of
F$_{H\beta}$ shown in the third row from the bottom of the table.
These line strengths in turn were corrected for reddening by assuming
that the relative strength of H$\alpha$/H$\beta$=2.86 and computing the
logarithmic extinction quantity $c$ shown in the penultimate line of
the table.  Values for the reddening coefficients,
f($\lambda$), are listed in column~(2), where we employed
Seaton's (1979) extinction curve for the UV and that of Savage \& Mathis
(1979) for the optical.

Because of imperfect spatial overlap between the optical and IUE
observations for all but DDDM-1, a final adjustment was made by
multiplying the IUE line strengths by a merging factor that was
determined from either the theoretical ratio of the He~II lines
$\lambda$1640/$\lambda$4686 or the carbon lines
C~III]~$\lambda$1909/C~II~$\lambda$4267.  The calculation of the
merging factors is described in Appendix A, and their values
are listed in the last row of Table~2.

The columns headed I($\lambda$) list our final, corrected line
strengths, again normalized to H$\beta$=100.  In general,
intensities have uncertainties $\le$10\%; single colons indicate
uncertainties up to $\sim$25\%, and double colons denote
doubtful detections with uncertainties up to $\ge$50\%.

\section{Results}

\subsection{Electron Temperatures And Densities}

Numerous temperature-sensitive line ratios are available in our data,
enabling us to sample electron temperatures at different positions
along the line-of-sight.  For example, [O~III] and [N~II]
temperatures, T$_{[O~III]}$ and T$_{[N~II]}$, are given by the intensity
ratios $\lambda$4363/($\lambda$4959+$\lambda$5007) and
$\lambda$5755/($\lambda$6548+$\lambda$6584), respectively.  Two other
temperatures are T$_{[O~II]}$ and T$_{[S~II]}$ which can be inferred
from the intensity ratios of $\lambda$7325/$\lambda$3727 and
$\lambda$4072/($\lambda$6716+$\lambda$6731), respectively.  In
addition, the intensity ratio of $\lambda$6716/$\lambda$6731 is
particularly sensitive to electron density, enabling the determination
of the [S~II] density, N$_{[S~II]}$.

Therefore, we have computed temperatures and densities for each
observed position, and the results are listed in Table~3.  For each
object listed in the first column we provide the temperatures and
densities in Kelvins and cm$^{-3}$, respectively, determined from the
above ratios. Values for T$_{[O~III]}$ and T$_{[N~II]}$ could be
calculated in all cases.  For T$_{[O~II]}$ and T$_{[S~II]}$, observed
line ratios often implied temperatures in excess of 25,000K, which
seemed unlikely to us, and thus we do not report values in those
cases.  A possible cause of these excessive temperatures could be the
tendency to overestimate the strengths of weak lines such as [O~II]
$\lambda$7325 and [S~II] $\lambda$4072.  Since electron temperatures
vary directly with these line strengths, overestimating them would
produce temperatures exceeding the actual values.

Based on our assessments of the errors in our line strength measurements,
we estimate the following uncertainties in our calculated temperatures
and densities: T$_{[O~III]}$ : $\pm$500K; T$_{[N~II]}$ : $\pm$1000K,
except for IC~3568, IC~4593, NGC~7009c, and all positions in NGC~6826,
which are $\pm$5000K; T$_{[O~II]}$: $\pm$2000K; T$_{[S~II]}$:
$\pm$4000K; N$_{[S~II]}$: $\pm$200 cm$^{-3}$, except for IC~3568,
which is $\pm$700 cm$^{-3}$.

We note that where multiple observations are available for different
positions within a single nebula, e.g. NGC~7009, NGC~6720, and
NGC~6826, values for the density or a specific temperature type are
quite consistent, particularly in the case of T$_{[O~III]}$.  Also
obvious is that the electron temperatures derived from both [O~II] and
[S~II] lines are consistently above those determined from [O~III] and
[N~II]. One possible explanation is the hardening of the radiation
field as it passes from the high-ionization zones nearer to the star
where [O~III] is found, out to the lower-ionization regions where
[O~II] and [S~II] are formed. However, one would think that [N~II]
should share this behavior, which it apparently does not.  Despite the
long baselines for both [O~II] and [S~II] measurements, reddening
errors are not to blame for the temperature discrepancy, since in the
case of [O~II] , the auroral lines are redward of the nebular lines,
whereas in [S~II], it is the nebular lines that are redder than the
transauroral lines.  Finally, we note again that weak line intensities
tend to be overestimated, which would result in electron temperatures
that are too high.

\subsection{Abundance Calculations}

In this project we are concerned with the abundance ratios of He/H,
O/H, C/O, N/O, and Ne/O. Paper~I describes our abundance calculation
method in detail. Future papers will use our newly acquired
optical data to study S/O and Ar/O.

Two distinct methods exist for deriving abundances in nebulae.  The
first employs measured line strengths for observed ions along with
knowledge of electron temperature and density to determine ionic
abundances using a set of simultaneous equations.  Subsequently, these
ionic abundances are converted to elemental abundances by using
standard correction factors to account for unobserved ions of an
element.  The second method involves the calculation of a detailed
photoionization model whose output line strengths match the observed
ones as closely as possible.  The input elemental abundances used to
produce the successful model are then taken to represent the true
levels in the real nebula.

Each of these methods has its own drawbacks.  In the first instance,
the correction factors can be sensitive to nebular properties such as
matter- or radiation-boundedness.  In the second case, the perennial
problem is determining the uniqueness of the model solution.  The
number of available observational constraints can be different for
each member of a sample such as ours.  Thus, a systematic approach
using only models to analyze a large number of objects suffers from
non-uniformity.

For these reasons we have developed a hybrid method in the spirit of
Shields et al. (1981).  The heart of the method is the use of a
photoionization model to improve results from a five-level atom
routine.  Briefly, for each PN (or for each position within a PN where
multiple positions were observed) we compile a set of merged UV and
optical line strengths and use the five-level atom routine ABUN to
derive an initial set of nebular abundance ratios A$^{PN}_{abun}$(X),
where X is one of the five abundance ratios listed above.  We then
employ the photoionization code CLOUDY version 84 (Ferland 1990) to construct a
nebular model, use ABUN to determine a set of model abundances
A$^{mod}_{output}$(X) based upon the model output line strengths, and compare
these with the actual model input abundance ratios
A$^{mod}_{input}$(X).  Our final set of abundance ratios
A$^{PN}_{F}$(X) for each PN (or each position within a PN) is
calculated by assuming that:

\begin{equation}
\eqnum{1a}
A^{PN}_{F}(X)=A^{PN}_{abun}(X) \xi(X)
\end{equation}
where
\begin{equation}
\eqnum{1b}
\xi(X)={{A^{mod}_{in}(X)}\over{A^{mod}_{abun}(X)}}.
\end{equation}

The correction factor $\xi$ is therefore a gauge of the accuracy
of the use of the ionization correction factor method for
determining abundances.  The program ABUN, including the sources
for atomic data, was described in detail in Paper~I.  Therefore,
we focus on the models used to determine $\xi$.

\subsection{Model Results}

Photoionization models were calculated for each slit position in order
to reproduce as closely as possible the physical conditions observed
along the line-of-sight.  Our models were constrained by a set of 10
important diagnostic ratios constructed directly from observed line
strengths.  These 10 ratios are known to describe the physical
conditions of a nebula quite well.  Our goal for each object (or
position within an object) was to match each observed ratio to within
0.10-0.15~dex, consistent with observational uncertainties.  We
assumed that the central stars were blackbodies and that the nebula
had a uniform density with a filling factor of unity.{\footnote{While
most PNe are known to have filling factors significantly less than
unity, the only measurable quantity affected by the filling factor is
the nebular luminosity, a parameter we are not using to constrain our
models.  Thus, using the same filling factor for all of the nebulae in
no significant way influences our abundance results.}}  The inner
nebular radius was taken to be 0.032~pc for all models, but the outer
radius was treated as a free parameter.  In several cases the best
matches to the observed line strengths were produced by truncating the
model inside the Str{\"o}mgren radius, i.e. the model nebula was
matter-bounded.  Other free parameters included the stellar
luminosity, nebular electron density, and nebular abundances of
helium, oxygen, nitrogen, carbon, neon, and sulfur.

Table~4A summarizes our model results; for each PN (or position within
a PN) we list logarithmically the observed and model-predicted values
for 10 important diagnostic line ratios in the upper section of the
table.  The first ratio is sensitive to gas-phase metallicity and
electron temperature, the second and third to the level of nebular
excitation, the fourth and fifth to electron temperature and density,
respectively, and the last five to abundance ratios in the order He/H,
N/O, S/O, C/O, and Ne/O.  The lower section of the table provides
important model input parameters: stellar effective temperature
(T$_{eff}$), the log of stellar luminosity log($L$),
electron density (N$_e$), and the inner and outer nebular
radii (R$_o$ and R; values of R which are less than the Str{\"o}mgren
distance, i.e. matter bounded models, are indicated with a
footnote). These are followed by six input abundance ratios.
(N.B. We emphasize that these 
abundance ratios are {\it not} our final abundances for
each object, but are the abundances necessary to produce the best
model.)

There are several important points about the model results that
require discussion.  First, we note that, with the exception of
DDDM-1, which is spatially unresolved, these models represent the
best-fit solutions to a specific line-of-sight position within a
nebula; they are {\it not} models of whole planetary nebulae.  Model
parameters were varied to match observed quantities.  Thus, the
stellar luminosity employed in a model may not be a dependable gauge
of the true value.  In the same way, physical size of a nebula is not
necessarily related directly to R$_o$ and R.  Second, the only major
discrepancy between observed and predicted values in Table~4A occurs
in IC~4593 for the He~II/He~I line ratio.  Large sections of parameter
space were explored in trying to render a match to this and the other
ratios observed for IC~4593.  We are encouraged, however, by the fact
that good matches have been achieved for the nine remaining ratios,
and suggest that the lack of agreement between theory and observations
for the He ratio is related to a peculiarity in the spectrum of the
real central star at the He$^+$ ionization edge.  Third, there is a
discrepancy of five orders of magnitude in the luminosities used to
model positions $a$ and $b$ in NGC~6720 (see Table~4A).  These values
were required to match the greatly different ratios observed for
log$I_{[O~II]}$/$I_{O~III]}$.  It may be that position~$b$ is heavily
shadowed and hence is characterized by much lower ionization, compared
with the situation at position~$a$.  Finally, we note that for those
PNe where more than one line-of-sight position was modelled, i.e.,
NGC~7009, NGC~6720, and NGC~6826, the input parameters for the
individual models agree quite closely, with the major differences
occurring for those parameters related to position, i.e. $L$, and R.

%
Table~4B lists the correction factors $\xi$ derived from the models,
where $\xi$ is the ratio of the input model abundance of an element to
the value derived from the model-predicted line strengths using the
program ABUN. A value of unity represents complete consistency between
the two abundance sets.  Therefore, $\xi$ is a model-determined gauge
of how closely the abundances derived with our 5-level atom program
agree with the actual nebular abundances.  An inspection of Table~4B
indicates that with some exceptions, most frequently for N/O, values
for $\xi$ are within 20\% of unity.  We have employed footnotes here
to indicate those models that are matter-bounded, and the large
discrepancies for N/O, all greater than unity, are clearly associated
with these models.  Since input N/O varies little from model to model
(see Table~4A), the large values of $\xi$ for N/O in these
matter-bounded models must be due to the way in which [N~II]
$\lambda$6584 relative to [O~II] $\lambda$3727 changes as one moves
outward approaching the Str{\"o}mgren edge.  Truncation results in a
quite different line ratio than would be predicted if the model were
to extend out to the Str{\"o}mgren radius.  This finding suggests that
N/O ratios in matter-bounded nebulae generally are less secure when
the standard ionization correction relation for this ratio is used.

\subsection{Derived Abundances}

Our final abundances for the seven PNe studied here are presented in
Table~5A and Fig.~1.  Results for each position, for those PNe where
more than one position was observed, are given along with averages.
Abundance values in Table~5A are given on a linear scale.  We point
out that our final abundances in Table~5A for any one object are very
consistent with the model input abundances given in Table~3A for the
same object.  Since the abundances in Table~5A are only indirectly
connected to the model abundances in Table~4A through the use of a
model-derived correction factor, this result is reassuring, albeit not
altogether surprising.  Our estimated uncertainties, not including
systematic effects, are 15\% for He/H, O/H, and Ne/O and 30-50\% for
C/O and N/O.  In Table~5A, the last row contains solar values for the
corresponding ratios taken from Grevesse \& Anders (1989) for
comparison, while the last column lists the Peimbert class of each
object.  Fig.~1 shows our abundance ratios in logarithmic form and
normalized to solar values, where averages have been plotted for those
PNe where more than one position was observed.  Ratios derived in this
paper are shown with filled symbols, using symbol shape to represent
specific objects as defined in the figure caption. For comparison,
abundance ratios taken from the literature (see the figure caption for
sources) are shown with open symbols.  Representative error bars are
given for each abundance ratio to show uncertainties.

Note that our results for He/H, O/H, and Ne/O are very consistent with
earlier measurements.  All seven PNe have sub-solar O/H, with the halo
object DDDM-1 being the most metal-poor, as expected.  In addition,
Ne/O is close to solar in all objects, consistent with the idea that
both of these elements are produced by stars of similar mass.

In the cases of C/O and N/O we see a large spread among our seven
objects.  Note that all objects have N/O above solar.  In addition,
C/O is markedly below solar in DDDM-1 and IC~4593 but well above solar
in NGC~6720 and NGC~7009.  In the case of IC~4593, our derived value
of C/O is roughly 20 times smaller than the ones published by
Bohigas \& Olgu{\'i}n (1996) and French (1983), both of whom used
optical recombination lines of carbon in their analysis.  Bohigas \&
Olgu{\'i}n derived their carbon abundance for IC~4593 using the C~III
$\lambda$4648 line, which is often blended with [O~II] $\lambda$4649,
while French used both C~III $\lambda$4648 and (uncertain) C~II
$\lambda$4267 to derive carbon abundances.  We note that our C/O value
for IC~4593 as estimated from C~II $\lambda$4267 in Table~5B is
consistent with our results from the C~III] $\lambda$1909 line.
Finally, our derived N/O values are higher for most objects, the
result of using values for $\xi$ which are significantly larger than
unity.

Straightforward determination of carbon abundances is hindered by the
longstanding problem whereby abundances inferred from the C~II
$\lambda$4267 recombination line are consistently several times greater than
values determined from the collisionally-excited line C~III]
$\lambda$1909.  (See Rola \& Stasi{\'n}ska 1994 for a recent discussion
of this problem.)  Since we were able to measure $\lambda$4267 in our
data at most observed locations, we have estimated C/O ratios implied
by this line, employing the data for the relevant effective
recombination coefficient given in P{\'e}quinot, Petitjean, \& Boisson
(1991).  Results of this exercise are shown in Table~5B, where for each
position listed in column~1 we give N(C$^{+2}$)$_{\lambda
4267}$/N(C$^{+2}$)$_{\lambda 1909}$, the predicted ratio of the number
density of C$^{+2}$, in column~2.  Then, since $C/O \approx
C^{+2}/O^{+2}$ (Rola \& Stasi{\'n}ska 1994; Paper~1), we scaled the C/O
ratios in Table~5A derived from the $\lambda$1909 line by multiplying
them by the values in column~2 of Table~5B to arrive at the estimates
of recombination C/O reported in column~3.  We give results for each
observed position along with an unweighted average for those PNe in
which we obtained data at more than one location.  Numbers in column~2
provide a good comparison of the recombination and collisional
excitation methods for inferring C/O.  As seen in many previous
studies, the recombination method consistently implies a significantly
larger value for C/O.  However, because of the uniformly weak strength
of C~II $\lambda$4267, we have adopted the C/O abundances determined
from the C~III] $\lambda$1909 line for the {\it final} values in our study
for this abundance ratio.

A discussion of implications for stellar nucleosynthesis is postponed
until abundances for our entire sample have been determined.

\section{Summary}

This paper is the third in a series reporting on a study of carbon
and other abundances in a well-defined sample of planetary nebulae
representing a broad range in progenitor mass and metallicity. 
We have obtained new optical spectra from 3700-9600{\AA} at
specific positions in our program objects in order to overlap
spatially as nearly as possible with earlier IUE
sites.  We 
make use of the Final Archive database of IUE spectra, in which the
data have been reduced under a new, consistent system of
algorithms. We have measured collisionally-excited emission lines of
carbon and coupled these measurements to our new optical line strengths
to determine abundance ratios of He/H, O/H, C/O, N/O, and
Ne/O for seven PNe: DDDM1, IC~3568, IC~4593, NGC~6210,
NGC~6720, NGC~6826, and NGC~7009.  
IUE and optical spectra for the same line-of-sight
position were merged using the ratio He~II
$\lambda$1640/$\lambda$4686, or the C~III] $\lambda$1909/C~II
$\lambda$4267 ratio when both He~II lines were unavailable.  In those
positions where both the helium and carbon lines were measurable, we
found reasonable consistency between these two methods.  To our
knowledge, this is the first attempt to use the carbon line ratio for
merging UV and optical spectra.

Electron temperatures and densities were determined at all locations.
Derived values for different positions within the same nebula were
found to be quite consistent.

Nebular abundances were inferred by using a hybrid method which
couples an empirical 5-level atom calculation with a photoionization
model which is used to fine-tune the ionization correction factor.
Thus, a photoionization model was produced for each object in which 10
important diagnostic line ratios were matched satisfactorily.  Many of
the models for our objects were matter bounded, and as a result
produced a large abundance correction factor for N/O in particular.
This finding implies that the standard ionization correction factors
for the N/O ratio are unsatisfactory when the nebula is optically
thin.

Our abundance results show a wide spread in C/O and N/O ratios among
the objects, while He/H, O/H, and Ne/O ranges are much narrower, all
of which is consistent with previous studies.  In those PNe
where more than one line-of-sight position was observed, derived
properties were consistent among those positions, thus adding credence
to our results and especially our abundance-determining techniques.
While our final C/O abundance ratios were determined using
collisionally-excited lines, we also estimated values for this ratio
using the recombination line C~II $\lambda$4267.  We found that ratios
produced by the latter method are consistently several times greater
than those produced by the former, as reported in many studies in the
literature.

\acknowledgments This project is supported by NASA grant NAG
5-2389. K.B.K. also acknowledges support from a Cottrell College
Science grant of Research Corporation and from the Bronfman Science
Center and the Dean of Faculty Office at Williams College.
R.B.C.H. is grateful to the University of Oklahoma for support of
travel to KPNO.  We thank Cathy Imhoff, Walter Feibelman and Fred
Bruhweiler for help tracking down IUE positional information. We thank
Jackie Milingo (OU) and Tim McConnochie (Williams '98) for their
assistance with the observations. We are grateful to KPNO for funding
McConnochie's observing trip. We also thank the ever-helpful KPNO
staff, especially Ed Carder and Jim De~Veny.

\appendix {A. Determination of the Merging Factor}

The optical and IUE observations were made using slits of different
sizes.  Due to this, as well as to inherent differences in the two
instruments and data-reduction algorithms, one needs to determine a
merging factor at each location, i.e. a number by which the UV line
strengths are multiplied in order to correct for a systematic offset
in the two flux-calibrated spectra for each nebular location.

Calculating a merging factor requires that we know both the
observed and theoretical values of a ratio involving two
emission lines, one appearing in the UV and the other in the
optical, but produced by the same ion.  The merging factor is
obtained by dividing the theoretical by the observed ratio.

Two such ratios are He~II $\lambda$1640/He~II $\lambda$4686 and C~III]
$\lambda$1909/C~II $\lambda$4267.  We measured the values of one or
both of these ratios in all of our target positions.  The theoretical
values for the He line ratio were determined using recombination
results in Storey \& Hummer (1995) along with necessary electron
temperatures and densities inferred from our optical observations.
The theoretical carbon line ratios were determined using the empirical
fit in Kaler (1986).  Because of the dependence of the C~III]/C~II
ratio on the generally weak C~II $\lambda$4267 recombination line as
well as the temperature-sensitive collisionally-excited C~III]
$\lambda$1909 line, the temperature-insensitive He ratio was preferred
for determining the merging factor.  Thus, when joining UV and optical
spectra, the carbon ratio was employed only in those cases when He~II
$\lambda$1640 was not observed.  Merging factors for NGC~7009,
NGC~6720, and NGC~6210 were computed from the helium lines, while for
IC~3568, IC~4593, and NGC~6826 the carbon lines were employed.

We note that for those objects for which both ratios were available,
we found good consistency between the two factors.  This is shown in
Fig.~A1, where we plot the merging factor determined from the helium
lines versus that from the carbon lines for those nebular positions
where both factors could be calculated.  The solid line shows where
points would lie if factors at each location were equal.  Clearly, the
He factor tends to be systematically larger than the C factor but for
reasons that are not obvious.  It is possible that this offset is
related to the tendency to overestimate the strength of a weak line
such as C~II $\lambda$4267.  However, there is good evidence here
overall for rough agreement in merging factors, a result which
supports the use of each of these factors.  We believe that this is
the first time that the carbon lines have been used in this manner to
combine UV and optical spectra.

The merging factor used to scale data at each location is given in the
last row of Table~2.  If the IUE and Goldcam slits had overlapped
perfectly, then the value of the merging factor should be unity,
ignoring instrumental effects.  Since the ratio of slit areas was 1:2
(Goldcam:IUE), then we would expect merging factors to fall within
roughly a factor of 2 of unity.  However, results in Table~2 show that
this number is below 0.3 for six of our 13 positions measured.  In
attempting to understand the cause of these low merging factors, we
considered such things as quality of slit overlap, as discussed in
{\S}2.3, the particular line ratio used to determine the merging
factor, excitation level of the object, electron density, and whether
or not the position appears to be matter-bounded.  For example,
although we believe the slit overlap for IC~3568 to be good, the
merging factor is unexpectedly low.  Likewise, the quality of slit
position match-up for the other five positions with merging factors
below 0.3 is ``reasonably good.''  In addition, for four of the six
positions in question, the merging factor was determined using the
carbon ratio, thus it is not obvious that the technique used to
determine the merging factor is itself at fault. Using the
[O~II]/[O~III] line ratio as an indicator of excitation, we find no
clear excitation difference between the six positions with low merging
factors and those with values closer to unity.  Nor does any pattern
emerge when either electron density or the question of
matter-boundedness is considered.  Curiously, however, all of the
merging factors differing significantly from unity are also {\it less
than} unity, indicating that fluxes measured with the IUE are
consistently too large with respect to the optical fluxes, even after
accounting for the factor-of-two difference in slit area.  While this
might be indicative of calibration problems with one of the
instruments, we would also expect the effect to appear at all other
positions as well, which is not the case.

Alas, there appears to be no obvious connection between these
properties and the calculated merging factor, and thus we are forced
to conclude that, while in most instances the slit overlap appears to
have been good, the smaller area of the Goldcam slit may have
frequently sampled regions with lower surface brightness, perhaps by
excluding knots present in the IUE slit.  However, we argue that given
the apparent consistency between the He and C merging factors shown in
Fig.~A1, their use in joining the UV and optical spectra for the
current objects is justified and is no doubt better than applying no
correction at all.

We plan to pursue the question of merging factors using a larger database in a future paper.

\clearpage


\figurenum{1} \figcaption{The five abundance ratios computed here and
normalized to their solar values (Grevesse \& Anders 1989) are plotted
logarithmically.  Symbol shapes are used to designate each object (see
legend).  Filled symbols are our results, while open symbols refer to
results from other studies found in the literature: values for IC~4593
were taken from Bohigas \& Olgu{\'i}n (1996), C/O for DDDM1 is from
Howard et al. (1996) and consistent with the upper limit given in
Clegg et al. (1987), all other C/O values are from Rola \&
Stasi{\'n}ska (1994; their ratios as derived from C~III]
$\lambda$1909), and all remaining abundance ratios are from Perinotto
(1991).}

\figurenum{A1} \figcaption{Helium merging factor versus carbon
merging factor for those nebular positions for which both could
be measured.  The solid line shows the locus of one-to-one correspondence.}

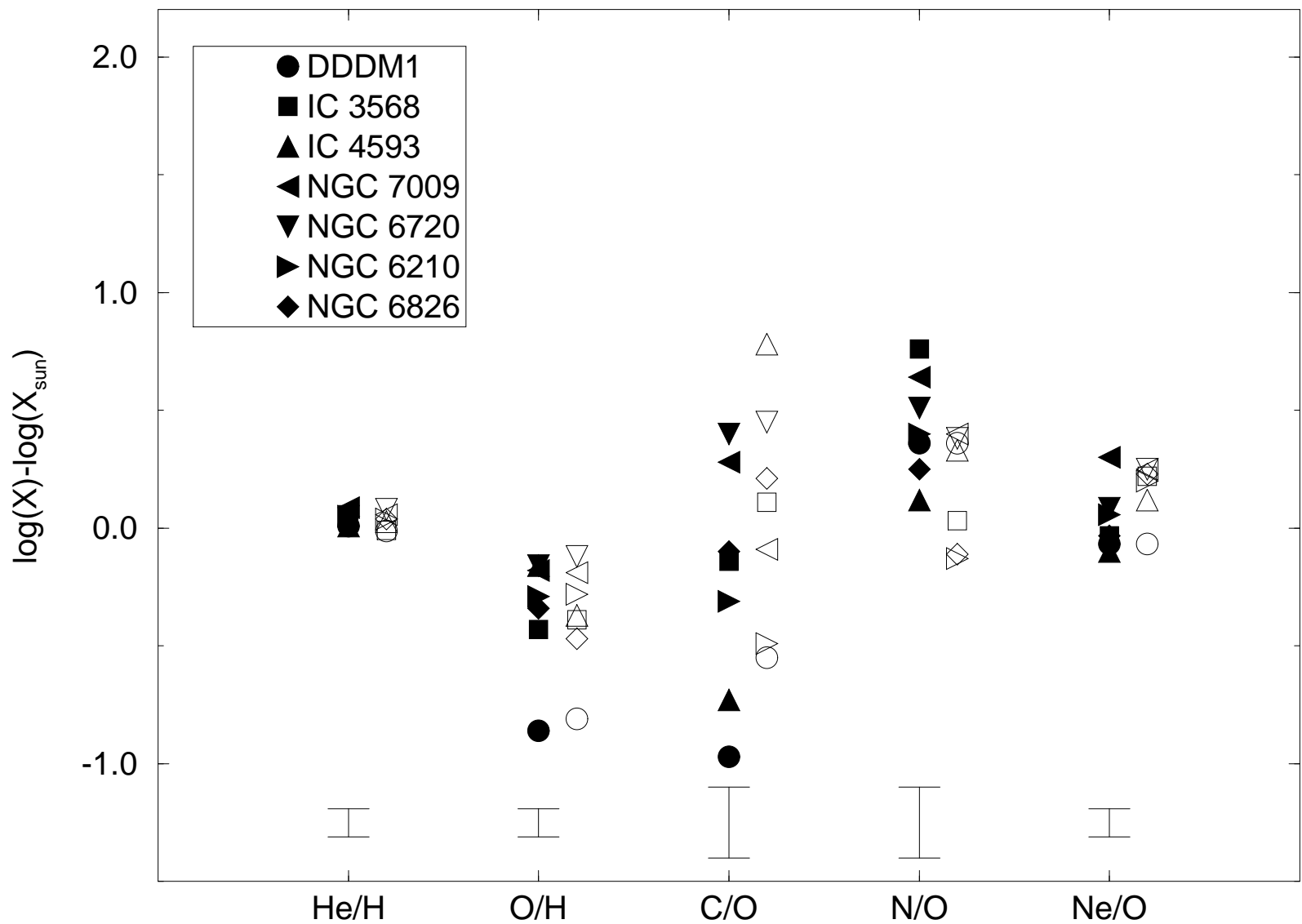
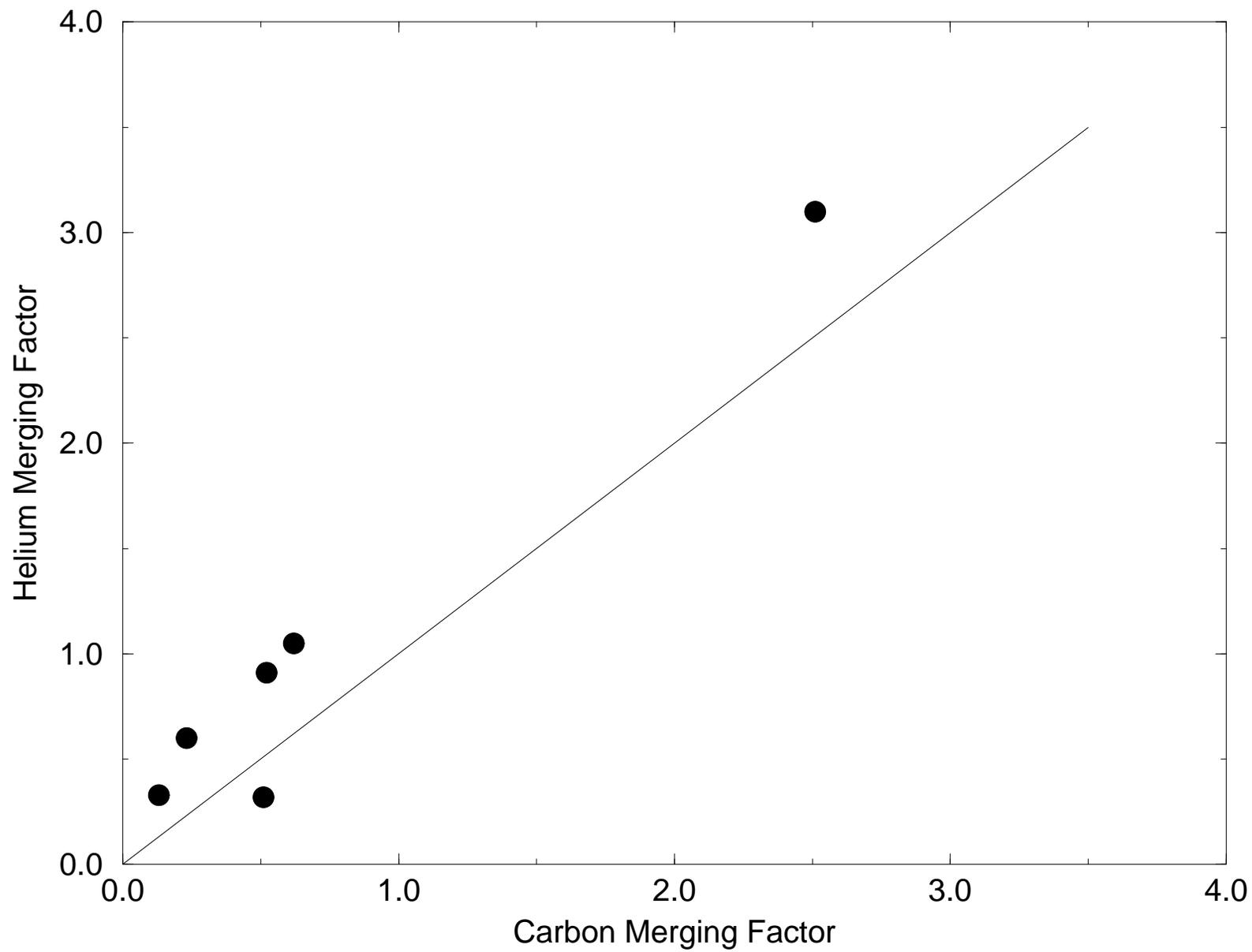
\end{document}